\documentclass[twocolumn,aps,showpacs,preprintnumbers,superscriptaddress,amsmath,amssymb,prl]{revtex4-1}
\usepackage{graphicx}
\usepackage{MnSymbol}
\usepackage{setspace}

\def\pr{Phys.\ Rev. }
\def\prep{Phys.\ Rep. }
\def\nl{Nano\ Lett. }

\def\opex{Opt.\ Express }
\def\sci{Science }
\def\cpl{Chem.\ Phys.\ Lett. }
\def\natm{Nature\ Mater. }
\def\natn{Nature\ Nanotechnol. }
\def\acid{Angew.\ Chem. }

\begin{document}
\title{Fano resonance resulting from a tunable interaction between molecular vibrational modes and a double-continuum of a plasmonic metamolecule}
\author{E. J. Osley}
\affiliation{London Centre for Nanotechnology, University College London, 17-19 Gordon Street,
London WC1H 0AH, UK} \affiliation{Department of Electronic and Electrical Engineering, University
College London, Torrington Place, London WC1E 7JE, UK}
\author{C. G. Biris}
\affiliation{Department of Electronic and Electrical Engineering, University College London,
Torrington Place, London WC1E 7JE, UK}
\author{P. G. Thompson}
\author{R. R. F. Jahromi}
\author{P. A. Warburton}
\affiliation{London Centre for Nanotechnology, University College London, 17-19 Gordon Street,
London WC1H 0AH, UK} \affiliation{Department of Electronic and Electrical Engineering, University
College London, Torrington Place, London WC1E 7JE, UK}
\author{N. C. Panoiu}
\affiliation{Department of Electronic and Electrical Engineering, University College London,
Torrington Place, London WC1E 7JE, UK}
\date{\today}

\begin{abstract}
Coupling between tuneable broadband modes of an array of plasmonic metamolecules and a vibrational
mode of carbonyl bond of poly(methyl methacrylate) is shown experimentally to produce a Fano
resonance, which can be tuned \textit{in situ} by varying the polarization of incident light. The
interaction between the plasmon modes and the molecular resonance is investigated using both
rigorous electromagnetic calculations and a quantum mechanical model describing the quantum
interference between a discrete state and two continua. The predictions of the quantum mechanical
model are in good agreement with the experimental data and provide an intuitive interpretation, at
the quantum level, of the plasmon-molecule coupling.
\end{abstract}
 \maketitle


The spectral control of optical absorption in metallic nanostructures has generated rapidly growing interest, both as a means for studying light-matter interaction at the deep-subwavelength scale and to develop new functional nanomaterials. Plasmonic materials provide an ideal platform for achieving enhanced optical interaction at the nanoscale as their primary building blocks consist of highly resonant particles, which strongly confine and enhance the optical field \cite{bde03n,zsm05pr,vme10rmp,m07book}. These properties are commonly employed to increase the optical coupling between plasmonic structures and an optically active medium, enabling many applications, including surface-enhanced Raman spectroscopy \cite{ne97s,kwk97prl}, surface-enhanced infrared absorption spectroscopy \cite{ep06apl,kln08cpl}, enhancement of the excitation rate and fluorescence of quantum dots (QDs) and molecules \cite{sas05nl,pmt06nn,abn06prl}, chemical sensing \cite{sa05nl,lmw10nl}, and biodetection \cite{ahl08nm,kln12nm}. One effective approach for tailoring the shape
of plasmonic resonances is the Fano interference between a narrow (discrete) resonance and a broad (continuous) one \cite{f61pr,zgw08prl,gm11prb,lzm10nm}.

A defining feature of Fano resonances is their pronounced spectral asymmetry. This can be exploited in applications that require high sensitivity to changes in the environment. Fano resonances in plasmonic systems can also provide physical insights into light-matter interaction at the nanoscale, as they are primarily the result of short-range, near-field interactions. Although Fano resonances in plasmonic systems have been chiefly investigated in a classical context, they can also be observed in quantum systems consisting of QDs, atoms, or molecules coupled to metallic nanoparticles \cite{ahl08nm,npc08prl,pak10nl,gfa11nl}. In this case the Fano resonance results from the interaction between a quantum discrete state and a classical continuum (localized or extended) plasmon mode. In many applications it is desirable to tune the strength of the interaction between the discrete and continuous state, and consequently dynamically change the shape of Fano resonances. However, since the materials and geometrical parameters of a plasmonic structure are difficult to control at the nanoscale, a convenient way to achieve optical tunability has not yet been attained.

\begin{figure}[t]\centering
\includegraphics[width=7cm]{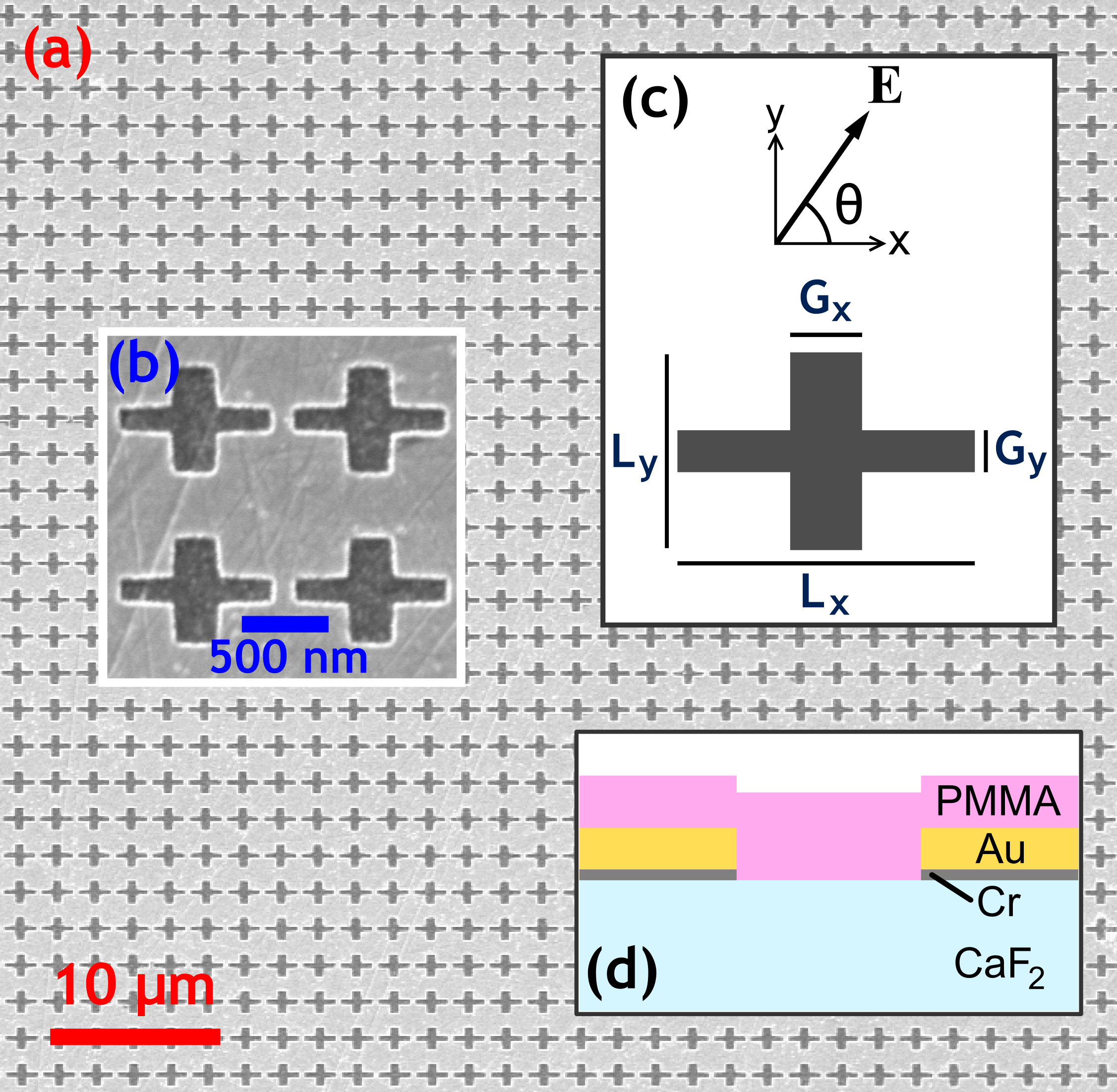}
\caption{(color online). (a) Scanning-electron micrograph showing part of the cruciform aperture
array before PMMA coating, (b) magnified detail of four apertures, (c) aperture geometry, showing the in-plane electric field polarization angle, $\theta$, (d) schematic cross section of the structure following PMMA coating.}
\label{fGeom}
\end{figure}

In this Letter we demonstrate experimentally and theoretically an efficient approach to
achieve \textit{in situ} tuning of the strength of the interaction between optically active
vibrational modes of molecules -- here the carbonyl bond of poly(methyl methacrylate)
(PMMA) -- and the localized surface plasmon (LSP) modes of plasmonic metamolecules consisting of
asymmetric cross apertures in a thin metallic film (see Fig. \ref{fGeom}). These nanostructures have been studied extensively \cite{rpa07ol,tbo11oe}, one of their defining properties being that, analogous to anisotropic molecules, their polarizability strongly depends on the polarization of the incident wave. These plasmonic metamolecules can therefore be tailored to different molecular resonances not only because the resonance frequencies of LSPs are strongly dependent upon their size and shape but, more importantly, because of their intrinsic optical anisotropy.

\textit{Sample fabrication}.---A $\mathrm{CaF}_{2}$ substrate of thickness $1~\textrm{mm}$ was
partially coated with a $5~\textrm{nm}$ chromium adhesion layer and a $35~\textrm{nm}$ gold film
by thermal evaporation. The array shown in Figs. \ref{fGeom}(a) and \ref{fGeom}(b) was
patterned using electron-beam lithography and argon-ion milling. The aperture geometry, and the
orientation of the electric field, $\mathbf{E}$, of incident polarized light relative to the
apertures is shown in Fig. \ref{fGeom}(c). The apertures have arm lengths $L_{x}=1900~\textrm{nm}$ and $L_{y}=1340~\textrm{nm}$ and arm widths $G_{x}=580~\textrm{nm}$ and $G_{y}=360~\textrm{nm}$. The surface was then spin-coated with a $85~\textrm{nm}$ PMMA film,
filling the apertures. Figure \ref{fGeom}(d) shows the final structure in cross section.

\textit{Optical characterization}.---The transmission and reflection spectra of the array were
measured using Fourier transform infrared spectroscopy (FTIR) microscopy. The source is a
broad-band mid-IR globar which is focused onto the sample. The light is collimated and passed through an aperture which selects radiation being transmitted or reflected from a circular sample area $79~\mathrm{\mu m}$ in diameter. The signal was detected using a HgCdTe detector. The reflection spectra $R$ and transmission spectra $T$ of the samples were normalized
to spectra obtained from uncoated areas of gold film and $\mathrm{CaF}_{2}$ substrate,
respectively.  The incident radiation was linearly polarized using a wire grid polarizer with an
extinction ratio of $168:1$ at $5~\mathrm{\mu m}$.
\begin{figure}[t]\centering
\includegraphics[width=8cm]{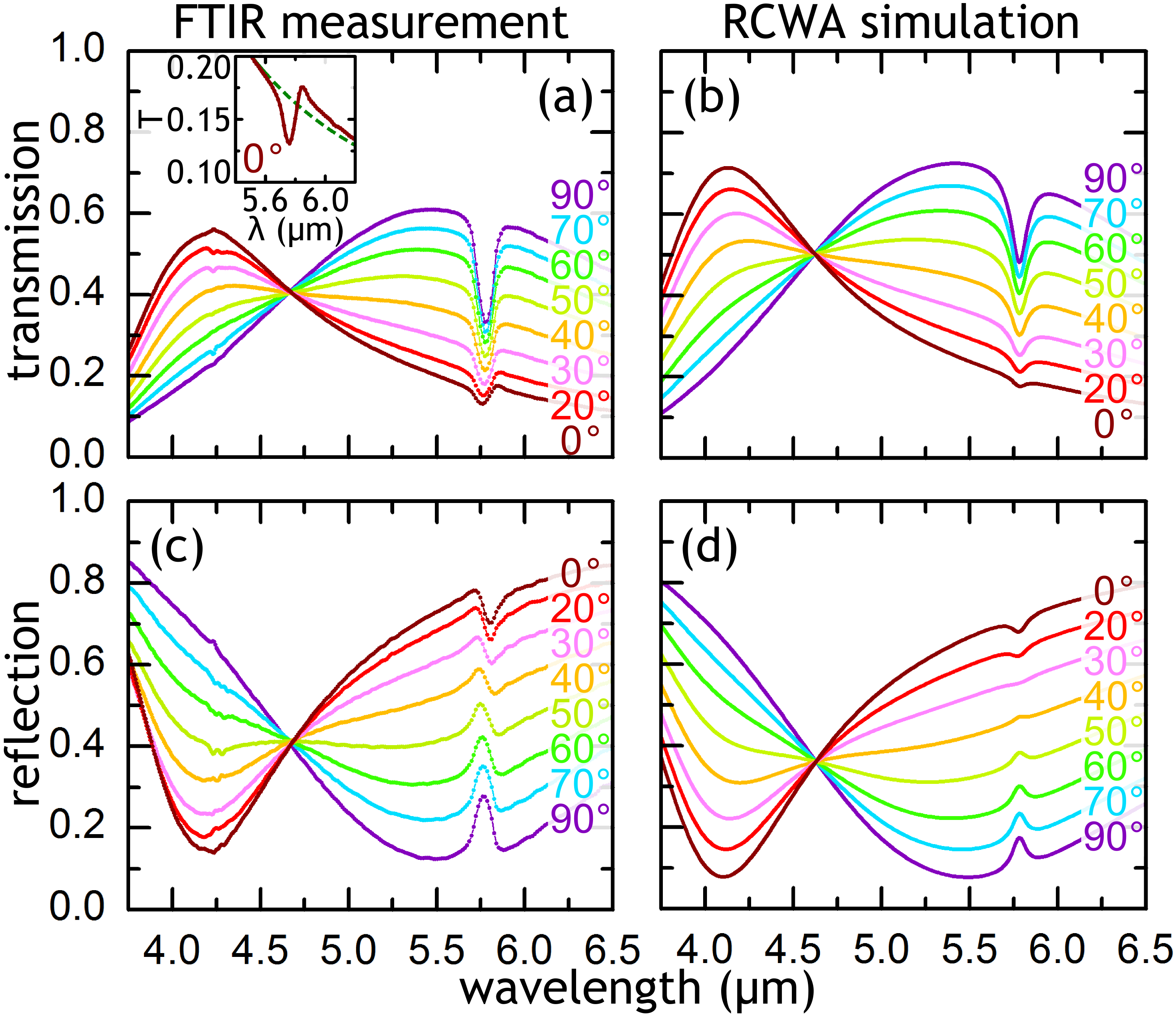}
\caption{(color online). (a) Transmission and (c) reflection of the structure measured
using FTIR; (b) transmission and (d) reflection simulated using the RCWA method. The incident polarization angle $\theta$ varies from zero (brown) to $\theta=90^{\circ}$ (purple). The resonance of the PMMA carbonyl bond is visible at $\lambda=5.79~\mathrm{\mu m}$. The inset to (a) shows the measured spectrum (solid line) and the fitted base-line (broken line) at the carbonyl bond resonance for $\theta=0^{\circ}$.
\label{fRe_Tr}}
\end{figure}

Figures \ref{fRe_Tr}(a) and \ref{fRe_Tr}(c) show, respectively, the measured transmission and reflection spectra of the aperture array coated with PMMA. In the transmission spectra two
global maxima are present; one, at $4.2~\mathrm{\mu m}$ occurs when the polarization angle $\theta=0$ and another, at $5.5~\mathrm{\mu m}$, when $\theta=90^{\circ}$. These global maxima are due to the excitation of two LSP modes, which are supported by the apertures \cite{rpa07ol,tbo11oe}. In the reflection spectra the excitation of the plasmon modes produces
global minima. The spectral signature of the PMMA carbonyl bond is located at $5.79~\mathrm{\mu
m}$. The shape of the carbonyl resonance peak changes with $\theta$, showing a more
pronounced, asymmetric Fano-like line shape as $\theta$ approaches zero. The interaction of the
two broad plasmon resonances with the narrow molecular resonance forms the Fano resonance. The inset to Fig. \ref{fRe_Tr}(a) also shows that, for small values of $\theta$, at wavelengths somewhat longer than the carbonyl bond resonant wavelength, the transmission can be larger than the baseline value in spite of the fact that PMMA is a lossy dielectric. This is an example of absorption-induced transparency [25]. Furthermore, as shown in Fig. \ref{fRe_Tr}(c) the reflectivity of the aperture array decreases with respect to the baseline for small $\theta$, whereas it increases for large $\theta$, by as much as $\sim$100$\%$ for $\theta = 90^{\circ}$. This phenomenon we refer to as \textit{absorption-induced reflectivity}. 

\begin{figure}[t]\centering
\includegraphics[width=7.5cm]{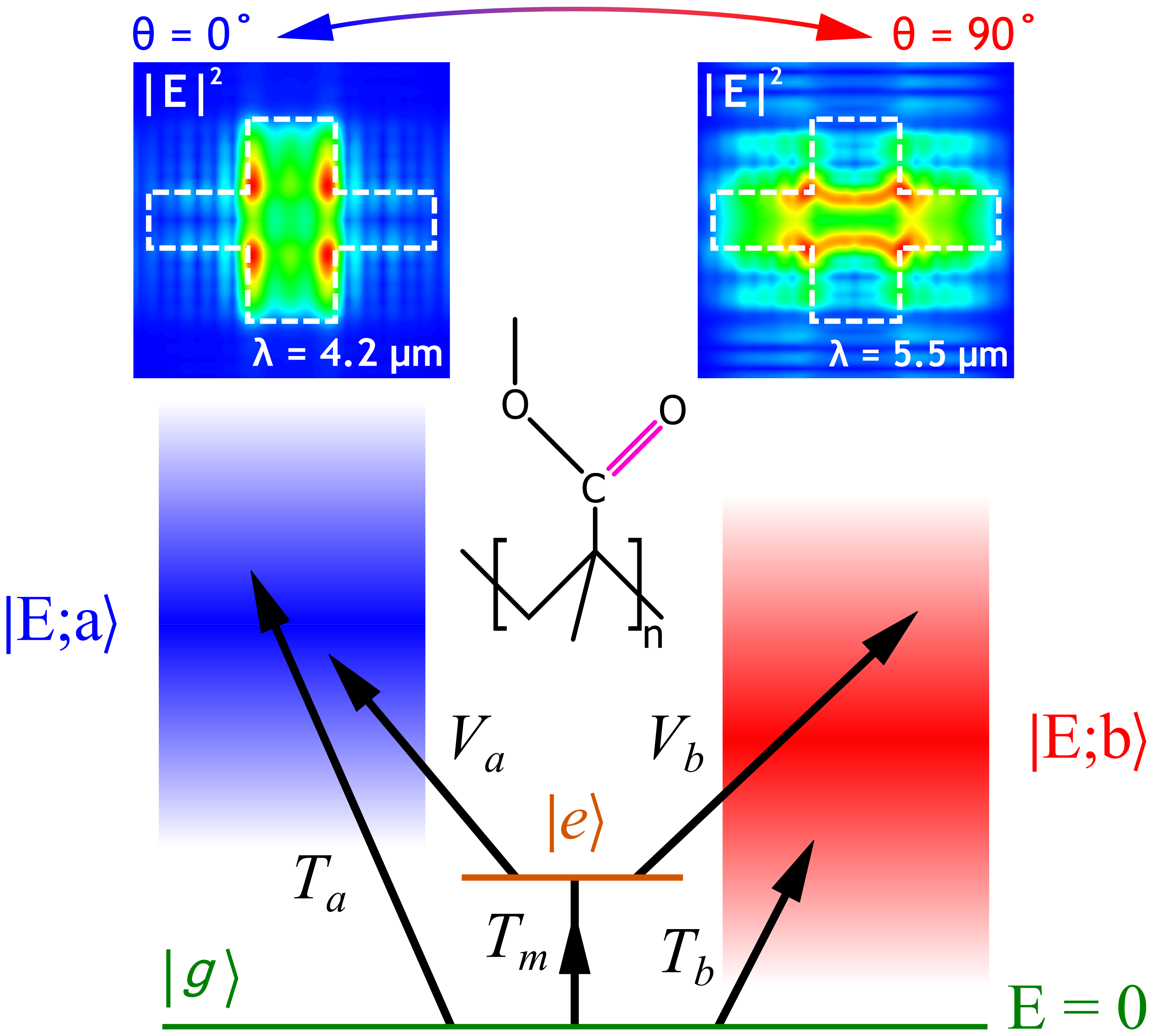}
\caption{(color online). System schematic showing the ground state $\vert g\rangle$, the two LSP
continua, $\vert E;a\rangle$ and $\vert E;b\rangle$, and the discrete molecular resonance $\vert
e\rangle$. Simulated spatial profiles (taken in the $x-y$ plane at the center of the aperture)
show large field enhancement for $\theta=0$ at $\lambda=4.2~\mathrm{\mu m}$ and
$\theta=90^{\circ}$ at $\lambda=5.5~\mathrm{\mu m}$. Aperture boundaries are indicated by dashed
lines. The atomic structure of the PMMA monomer is shown with the carbonyl bond in pink; the
resonance of this bond corresponds to the excited state $\vert e\rangle$. \label{QMcontinuum}}
\end{figure}

\textit{Numerical simulations}.---To model the optical response of our PMMA-coated device we used commercially available software \cite{RSoft} based on the rigorous coupled wave analysis (RCWA). We used a constant refractive index for the $\mathrm{CaF}_{2}$ substrate, a Lorentz-Drude model for the Au and Cr layers \cite{oba85ao}, and a complex refractive index for
the PMMA layer (found by fitting the numerically determined transmission to an experimental
measurement of the transmission of a $85~\mathrm{nm}$ thick PMMA film). Figures \ref{fRe_Tr}(b)
and \ref{fRe_Tr}(d) show the transmission and reflection spectra, respectively, found by numerical simulations; there is good agreement with the experimental data. The simulations also allow spatial field profiles to be calculated; Fig. \ref{QMcontinuum} shows the spatial profile of
the field enhancement corresponding to the two LSPs.

\textit{Theoretical model for molecule-plasmon metamolecule interaction}.---In what follows we
introduce a simple quantum mechanical model describing the interaction between the molecular
vibrational mode and the LSP resonances. Figure \ref{QMcontinuum} shows the interacting quantum
mechanical system schematically.

We assume that the interacting system is described by a Hamiltonian,
$\hat{H}=\hat{H}_{0}+\hat{V}_{a}+\hat{V}_{b}$, where
$\hat{H}_{0}=\hat{H}_{m}+\hat{H}_{a}+\hat{H}_{b}$ is the Hamiltonian of the noninteracting
molecule ($\hat{H}_{m}$) and the two LSPs ($\hat{H}_{a}$ and $\hat{H}_{b}$) and $\hat{V}_{a}$ and $\hat{V}_{b}$ describe the interaction between the molecule and the two LSPs. The molecule is modeled as a two-level system with $\vert g\rangle$ and $\vert e\rangle$ the ground and excited states, respectively. We assume that the eigenstates corresponding to the energy $E$, $\vert E;a\rangle$ ($\vert E;b\rangle$), of $\hat{H}_{a}$ ($\hat{H}_{b}$) form two continua. Note that the two plasmon modes are orthogonal and thus the continua are orthogonal, too: $\langle E;a\vert E';b\rangle=\delta_{ab}\delta(E-E')$.

Under these conditions we can write $\hat{H}_{0}$ as
\begin{equation}\label{nintHam}
    \hat{H}_{0}=E_{g}\vert g\rangle\langle g\vert+E_{e}\vert e\rangle\langle e\vert+\sum_{i=a,b}\int E\vert E;i\rangle\langle
    E;i\vert dE,
\end{equation}
where $E_{g}$ and $E_{e}$ are the energies of the ground and excited state of the molecule,
respectively. The interaction between the molecule and the two LSPs (continua) can be expressed as
$\hat{V}_{i}=\int V_{i,E}\vert E;i\rangle\langle e\vert dE+\mathrm{h.~c.}, i=a,b$, where
$V_{i,E}=\langle E;i\vert \hat{V}_{i}\vert e\rangle, i=a,b$, are the matrix elements of
$\hat{V}_{i}$ and are proportional to the local electric field of the LSP and $\mathrm{h.~c.}$ means Hermitian conjugate. Introducing the density of states of
the two continua, $\rho_{a,b}(E)$, these matrix elements are written as
$V_{i,E}=v_{i}\sqrt{\rho_{i}(E)}, i=a,b$. We assume that the spectra of the
plasmon \textit{A} (\textit{B}) is described by a Lorentzian centered at the plasmon
energy $E_{a}$ ($E_{b}$) with width $\Gamma_{a}$ ($\Gamma_{b}$),
\begin{equation}\label{Lf}
   \rho_{i}(E)=\frac{1}{1+{\left[\frac{2\left(E-E_{i}\right)}{\Gamma_{i}}\right]}^{2}},~~~i=a,b.
\end{equation}
Following the approach of Fano \cite{f61pr}, the total Hamiltonian $\hat{H}$ can be
orthogonalized in a new basis of eigenstates which form two orthogonal continua, \textit{i.e.},
$\langle E;1\vert \hat{H}\vert E';1\rangle=\langle E;2\vert \hat{H}\vert
E';2\rangle=E\delta(E-E')$, where $\langle E;1\vert E';2\rangle=0$. These eigenstates can be
written as \cite{f61pr}
\begin{equation}\label{cont}
  \vert E;1\rangle = \frac{\sin \Delta}{\sqrt{\pi\gamma}}\vert \Phi_{E}\rangle - \cos \Delta\vert \overline{\psi}_{E}\rangle;~~~\vert E;2\rangle = \vert
  \widetilde{\psi}_{E}\rangle,
\end{equation}
where $\vert \Phi_{E}\rangle$ is the ``dressed'' excited state $\vert e\rangle$ modified by the
interaction with the two continua,
\begin{equation}\label{discrete}
    \vert \Phi_{E}\rangle=\vert e\rangle+\sum_{i=a,b}\strokedint dE'\frac{V_{i,E'}}{\sqrt{\pi\gamma(E')}(E-E')}\vert
    E';i\rangle,
\end{equation}
(here $\strokedint$ means the Cauchy principal value of the integral) and $\vert
\overline{\psi}_{E}\rangle$ and $\vert \widetilde{\psi}_{E}\rangle$ are two continuum states
defined as:
\begin{subequations}\label{conts}
\begin{eqnarray}
  \label{c1s}\vert \overline{\psi}_{E}\rangle &=& \sqrt{\pi/ \gamma(E)}\left(V_{a,E}\vert E;a\rangle+V_{b,E}\vert E;b\rangle\right), \\
  \label{c2s}\vert \widetilde{\psi}_{E}\rangle &=& \sqrt{\pi/\gamma(E)}\left(V_{b,E}^{*}\vert E;a\rangle+V_{a,E}^{*}\vert
  E;b\rangle\right).
\end{eqnarray}
\end{subequations}
In Eq. (\ref{cont}) $\Delta$ determines the phase shift induced by quantum interference between
the discrete state $\vert e\rangle$ and the two continua, and $\gamma$ describes the width of the
modified excited state. They are given by:
\begin{subequations}\label{DG}
\begin{eqnarray}
  \label{D}\gamma(E) &=& \pi\left(\vert V_{a,E}\vert^{2}+\vert V_{b,E}\vert^{2}\right)\equiv\frac{\Gamma_{m}}{2}, \\
  \label{G}\Delta(E) &=& -\arctan\frac{\gamma(E)}{E-E_{e}-\Sigma(E)},
\end{eqnarray}
\end{subequations}
where the energy shift of the excited state is
\begin{equation}\label{sigma}
    \Sigma(E)=\strokedint \frac{\gamma(E')}{\pi(E-E')}dE'.
\end{equation}

The interaction between the coupled molecule-metamolecule system and a plane
wave normally incident onto the array can be described by a Hamiltonian
$\hat{T}=\hat{T}_{m}+\hat{T}_{a}+\hat{T}_{b}$, with $\hat{T}=t_{e0}\vert e\rangle\langle
0\vert+\sum_{i=a,b}\int T_{i,E}\vert E;i\rangle\langle 0\vert dE+\mathrm{h.~c.}$. Here, $\vert
0\rangle$ is the ground state of the interacting system, in which the molecule is in the ground
state and no plasmons are excited. The matrix elements of $T$ are: $t_{e0}=\langle
e\vert\hat{T}_{m}\vert 0\rangle$, $T_{a,E}=\langle E;a\vert\hat{T}_{a}\vert
0\rangle\equiv\mu_{a}\sqrt{\rho_{a}(E)}\sin \theta,$ and $T_{b,E}=\langle E;b\vert\hat{T}_{b}\vert
0\rangle\equiv\mu_{b}\sqrt{\rho_{b}(E)}\cos \theta$, where $\mu_{a,b}$ quantify the coupling
strength between the incoming plane wave and the two LSPs and we have used the fact that the
plasmon \textit{A} (\textit{B}) is excited by a \textit{y}(\textit{x})-polarized wave.

Using Fermi's golden rule, the rate at which the system absorbs photons of frequency $\omega$ is
$\mathcal{W}(\omega)=\frac{2\pi}{\hbar}\sum_{i=1,2}\int dE\vert\langle E;i\vert\hat{T}\vert
0\rangle\vert^{2}\delta(\hbar\omega-E)$. Inserting Eqs. (\ref{cont}) in this equation one obtains the absorption of the system,
\begin{equation}\label{abs}
    A(\omega)=\mathcal{N}\omega\left[\vert\langle\overline{\psi}_{\hbar\omega}\vert\hat{T}\vert 0\rangle\vert^{2}F(q,\epsilon)+\vert\langle\widetilde{\psi}_{\hbar\omega}\vert\hat{T}\vert
    0\rangle\vert^{2}\right],
\end{equation}
where $\mathcal{N}=2\pi n_{s}/I_{0}$ is a constant, with $n_{s}$ and $I_{0}$ being the number of
molecules per area and the intensity of the incoming beam, respectively, and
$F(q,\epsilon)=(q+\epsilon)^{2}/(\epsilon^{2}+1)$ is the Fano function. The asymmetry parameter,
$q$, and the reduced energy, $\epsilon$, are given by
\begin{equation}\label{Fp}
    q=\frac{\langle\Phi_{\hbar\omega}\vert\hat{T}\vert 0\rangle}{\sqrt{\pi\gamma(\hbar\omega)}\langle\overline{\psi}_{\hbar\omega}\vert\hat{T}\vert
    0\rangle};~\epsilon=\frac{\hbar\omega-E_{e}-\Sigma(\hbar\omega)}{\gamma(\hbar\omega)}.
\end{equation}

The absorption spectrum in Eq. (\ref{abs}) shows that the coupling of the molecular vibrational
mode with the two continua of the plasmonic metamolecule leads to a resonant response of the
optical system around the energy of the vibrational mode, $\hbar\omega\simeq E_{e}$, which is
superimposed on a broad background. Hence, unlike the case of only one continuum, the absorption
does not cancel at the frequency for which $F(q,\epsilon)=0$, \textit{i.e.}, when $q=-\epsilon$.

The interaction between the incoming light and the plasmons can be tuned by varying the polarization angle; this allows one to modify the mixture of the two LSPs and, consequently, one can easily tune the strength of the interaction between the vibrational mode and
the plasmonic metamolecule. In order to provide a more quantitative description we assume that the normalized densities of state of the two LSPs are given by Eq. (\ref{Lf}). Then, the matrix elements in Eq. (\ref{abs}) and the parameters of the Fano function can be calculated explicitly. The absorption can be cast as
\begin{align}\label{absf}
    A(\omega)=\frac{\mathcal{N}\pi\omega}{\gamma(\hbar\omega)}&\left[\vert \mu_{a}v_{a}^{*}\rho_{a}\sin\theta+\mu_{b}v_{b}^{*}\rho_{b}\cos\theta\vert^{2}F(q,\epsilon)+\right. \nonumber \\
    &\left. \rho_{a}\rho_{b}\vert\mu_{a}v_{b}\sin\theta+\mu_{b}v_{a}\cos\theta\vert^{2}\right],
\end{align}
where the Fano parameters are
\begin{subequations}\label{Fpf}
\begin{eqnarray}
    \label{qf}q &=& 2\frac{\frac{t_{e0}}{2\pi}+\frac{\hbar\omega-E_{a}}{\Gamma_{a}}\mu_{a}v_{a}^{*}\rho_{a}\sin\theta+\frac{\hbar\omega-E_{b}}{\Gamma_{b}}\mu_{b}v_{b}^{*}\rho_{b}\cos\theta}{\mu_{a}v_{a}^{*}\rho_{a}\sin\theta+\mu_{b}v_{b}^{*}\rho_{b}\cos\theta},~ \\
    \label{epsf}\epsilon &=& \frac{2}{\Gamma_{m}}\left[\hbar\omega-E_{e}-\pi\left(\rho_{a}\vert v_{a}\vert^{2}\frac{\hbar\omega-E_{a}}{\Gamma_{a}}+\rho_{b}\vert
    v_{b}\vert^{2}\frac{\hbar\omega-E_{b}}{\Gamma_{b}}\right)\right]. \nonumber \\
\end{eqnarray}
\end{subequations}
\begin{figure}[t]\centering
\includegraphics[width=7cm]{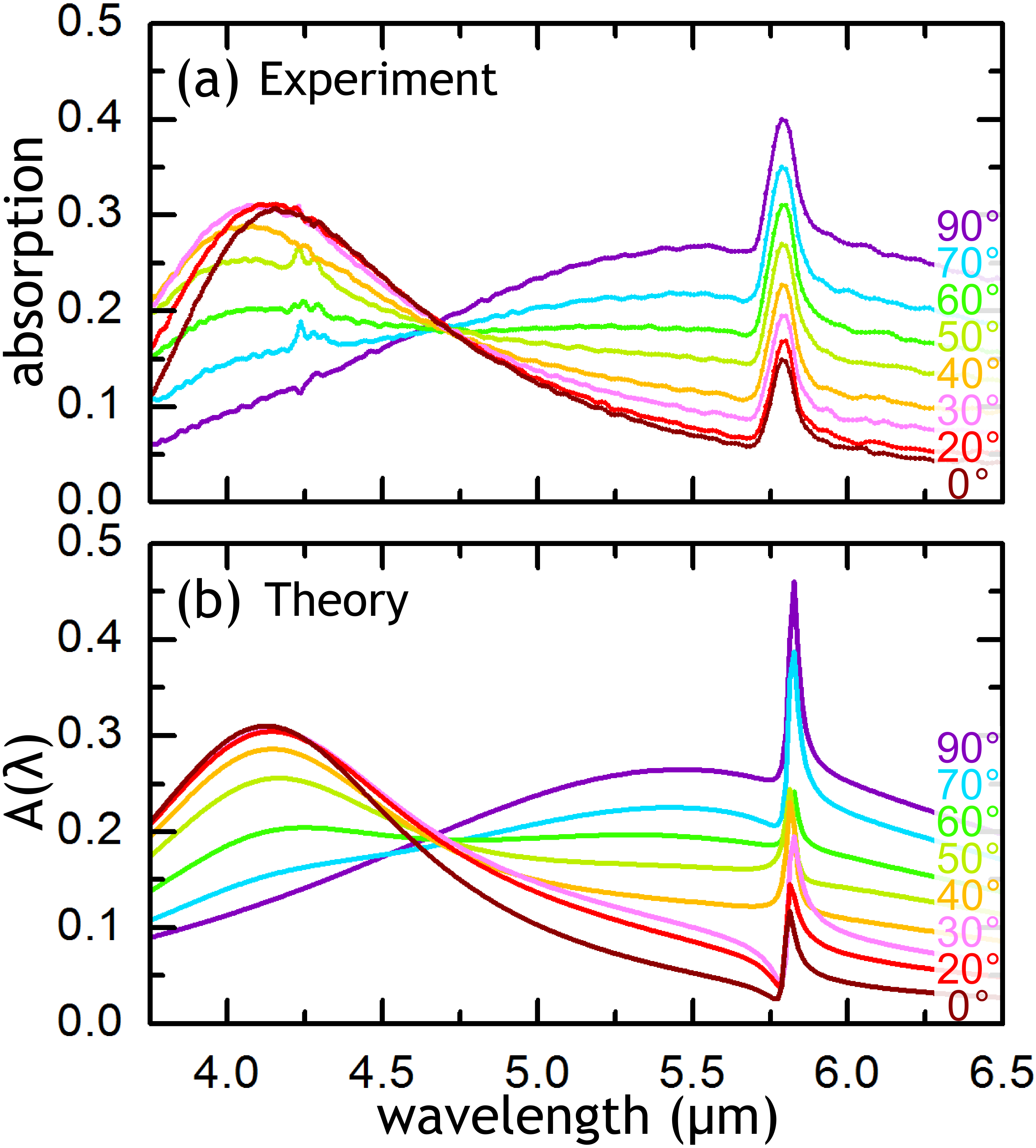}
\caption{(color online). (a) Absorption of the structure, from FTIR measurements of
transmission and reflection. (b) Best fit of experimental data using Eq. (\ref{absf}).
\label{QMoutput}}
\end{figure}

\textit{Comparison between theory and experiment}.---To fit the model to the experimental system the absorption $A$ of the structure was extracted from the measured transmission and reflection data using the relation $A=1-T-R$, as shown in Fig. \ref{QMoutput}(a). From this the parameters $E_{a,b}$ and $\Gamma _{a,b}$ can be directly extracted; $E_{a}$ and $\Gamma_{a}$ from the $\theta = 0$ data and $E_{b}$ and $\Gamma_{b}$ from the $\theta = 90^{\circ}$ data. $E_{e}$ and $\Gamma_{m}$ (the carbonyl bond parameters) are taken from transmission measurements of a 100 nm PMMA film. The parameters $v_{a,b}$, $\mu_{a,b}$, and $t_{e0}$ are then extracted for each value of $\theta$ using an unconstrained least squares method.

Figure \ref{QMoutput}(b) shows the predictions of the model following fitting. There is good
agreement between the measured spectra and those predicted by the model at all values of $\theta$. In particular, all the main features of the Fano resonance are reproduced by our theoretical model, \textit{e.g.} the increase of the optical absorption as $\theta$ is varied from zero to $90^{\circ}$. Our model also predicts that $t_{e0}\ll\vert v_{a,b}\vert^{2}$ for all angles $\theta$, which means that, as expected, the molecules interact primarily with the optical near-field of the SPPs. This proves that our model can be used to gain insights into the interaction between quantum and plasmonic systems as it allows one to retrieve the values of basic quantities, \textit{e.g.} $v_{a,b}$, $\mu_{a,b}$, and $t_{e0}$, which characterize these interactions at the nano-scale.

In conclusion, we have demonstrated that plasmonic structures allow \textit{in situ} tuning of the optical coupling between a quantum and a classical system simply by selecting the polarization of incident light. This tuning produces a dramatic change in the spectral line shape about the optical resonance of the quantum system, a consequence of the Fano resonance produced by the coupling of broad and narrow resonances. The system has been described using an intuitive quantum mechanical model, which allows a deeper understanding of the underlying physics. Our study has the potential to foster exciting new developments in sub-wavelength optics and nanophotonics and in the emerging field of quantum plasmonics. Equally important, the system investigated here has the potential to improve the range and versatility of a series of nanodevices, including chemical and biological plasmonic sensors.

This work was supported by EPSRC.

\end{document}